\documentclass[conference]{IEEEtran}

\usepackage{cite}
\usepackage{amsmath,amssymb,amsfonts}
\usepackage{algorithmic}
\usepackage{graphicx}
\usepackage{textcomp}
\usepackage[dvipsnames, table]{xcolor}
\usepackage{babel}
\usepackage{xspace}
\usepackage[capitalise,nameinlink]{cleveref}
\usepackage[inline]{enumitem}
\usepackage{booktabs}
\usepackage[nolist]{acronym}
\usepackage[
  babel=true,
  expansion=alltext,
  protrusion=alltext-nott,
  final
]{microtype}

\makeatletter
\@ifundefined{missingfigure}{}{}
\@ifundefined{textcomment}{}{}
\@ifundefined{sidecomment}{}{}
\@ifundefined{todo}{}{}
\@ifundefined{TODO}{}{}
\@ifundefined{todofix}{}{}
\@ifundefined{change}{}{}
\makeatother

\newcommand{\ie}{i.e.,}
\newcommand{\rscore}{$R^2$ }

\begin{acronym}
\acro{AI-NDT}{AI-based network digital twin}
\acro{DT}{digital twin}
\acro{GNN}{graph neural network}
\acro{IoT}{internet of things}
\acro{KG}{knowledge graph}
\acro{LSTM}{long short-term memory}
\acro{MAD}{median absolute deviation}
\acro{MAE}{mean absolute error}
\acro{ML}{machine learning}
\acro{MOS}{mean opinion score}
\acro{NDT}[NDT]{network digital twin}
\acro{NTKG}{network topology knowledge graph}
\acro{NSKG}{network state knowledge graph}
\acro{ASKG}{application state knowledge graph}
\acro{RMSE}{root mean square error}
\acro{RTT}{round trip time}
\acro{QoE}{quality of experience}
\acro{QoS}{quality of service}
\end{acronym}

\begin{document}

\title{On Effectiveness of Graph Neural Network Architectures for Network Digital Twins (NDTs)}

\author{
  \IEEEauthorblockN{Iulisloi Zacarias, Oussama Ben Taarit, Admela Jukan}
  \IEEEauthorblockA{Technische Universit\"at Braunschweig\\
   Braunschweig, Germany\\
   Email: \{i.zacarias, o.ben-taarit, a.jukan\}@tu-braunschweig.de}
}

\maketitle

\begin{abstract}
Future networks, such as 6G, will need to support a vast and diverse range of interconnected devices and applications, each with its own set of requirements. While traditional network management approaches will suffice, an automated solutions are becoming a must. However, network automation frameworks are prone to errors, and often they employ ML-based techniques that require training to learn how the network can be optimized. In this sense, network digital twins are a useful tool that allows for the simulation, testing, and training of AI models without affecting the real-world networks and users. This paper presents an AI-based Network Digital Twin (AI-NDT) that leverages a multi-layered knowledge graph architecture and graph neural networks to predict network metrics that directly affect the quality of experience of users. An evaluation of the four most prominent Graph Neural Networks (GNN) architectures was conducted to assess their effectiveness in developing network digital twins. We trained the digital twin on publicly available  measurement data from RIPE Atlas, therefore obtaining results close to what is expected in real-world applications. The results show that among the four architectures evaluated, GraphTransformer presents the best performance. However, other architectures might fit better in scenarios where shorter training time is important, while also delivering acceptable results. The results of this work are indicative of what might become common practice for proactive network management, offering a scalable and accurate solution aligned with the requirements of the next-generation networks.
\end{abstract}



\section{Introduction}
\label{sec:introduction}

\Acp{NDT} play an increasingly important role in network management, complementing automation frameworks by mirroring the physical network characteristics and enabling ``what-if'' analysis, testing configurations and algorithms, and training machine learning models for deployment in the network without impacting live users and applications ~\cite{masaracchia2022}. The use of \ac{NDT} significantly reduces the potential for service disruptions and offers comprehensive visibility of the network while minimizing costs associated with testing various scenarios. Today, different techniques can be employed to create the \ac{NDT}, with network simulators and emulators being the most popular ones. While \acp{NDT} offers significant advantages, their effectiveness is often hampered by the high computational cost of running those simulation tools and the challenge of accurately modeling complex, dynamic network behaviors~\cite{galmes2022}. \Ac{ML} and, particularly, \acp{GNN} offer a powerful alternative for modeling graph-structured data~\cite{chang2022}, enabling the extraction of insights and predictive analytics for the network system. 

As of today, modeling and such complex systems is still an open issue, and multiple alternatives in the realm of \ac{ML} are available. How to model the different components of a computer network (applications, network state, topology)  is also an open issue in the context of the key performance metrics predictions that directly affect users' quality of experience~\cite{schwarzmann2022}. While recent work proposed \ac{GNN}-based architectural designs 
\cite{chang2022, zhu2024, raj2023}, an evaluation of the  most effective for modeling real-world network behavior and predicting performance metrics, like \ac{RTT} and packet loss, is still missing. Ultimately, a trade-off between predictive accuracy, training efficiency, and model complexity among different GNN architectures needs to be evaluated, ideally when applied to a real-world network dataset.

This work presents an \ac{NDT} model that leverages \ac{GNN} for efficiently estimating key network performance metrics, as they strongly relate to the \ac{QoE} indicators in networks. In other words, the proposed \ac{NDT} can be used by network administrators or automation frameworks to estimate user satisfaction with a networked application or service. We propose an \ac{NDT} framework that integrates a multi-layered \ac{KG} architecture and provides a comparative analysis of four \ac{GNN} models applied to it.  We trained the digital twin on publicly available  measurement data from RIPE Atlas\cite{ripe-atlas}, therefore obtaining results close to what is expected in real-world applications. The results show that among the four architectures evaluated, GraphTransformer presents the best performance. However, other architectures may be better suited for scenarios where shorter training time is important, while also delivering acceptable results. The results of this work are indicative of what might become standard practice for proactive network management, offering a scalable and accurate solution aligned with the requirements of next-generation networks.

The remainder of this paper is organized as follows: \cref{sec:related-work} presents the background and discusses existing solutions and related work. \cref{sec:architecture} presents the architecture of the digital twin and describes each of its components. \cref{sec:results} presents the evaluation of the implemented models and discusses the main findings. Finally, \cref{sec:conclusions} concludes the paper, outlining future directions.




\section{Related Work}
\label{sec:related-work}

Michael Grieves was one of the first to introduce the concept of \ac{DT} in \cite{grieves2016}. He defined it  as ``a set of virtual information constructs that fully describes a potential or actual physical manufactured product from the microatomic level to the macrogeometric level.'' Its application to computer networks is recently gaining traction due to its ability to create virtual replicas of a physical network that can be used to simulate, test, and track network performance without impacting users utilizing the physical network infrastructure and without requiring physical hardware~\cite{qin2024}. 

Emulators and simulators have been among the most popular techniques to construct \acp{NDT}. However, these traditional methods face substantial drawbacks in terms of scalability. For emulation-based techniques, the \ac{NDT} must run a virtual instance of the devices and usually emulate different hardware architectures, which incurs a high computational cost. Simulation-based techniques are powerful and precise in modeling individual aspects of networks, like queuing. 
However, those models quickly get complex as more features are included to replicate the behavior of the whole system~\cite{galmes2022}. Modern research continues to build on simulation and emulation principles, as seen in the B5Gemini~\cite{Mozo2022}, an \ac{NDT} architecture for 5G and beyond, designed to emulate complex 5G scenarios with real-time synchronization capabilities. Another example is a simulation-based \ac{NDT} developed to train ML models and improve transmission quality in optical networks~\cite{shen2024}.

Recent work has explored integrating \ac{ML} to create more adaptive and predictive \acp{NDT}. \Acp{GNN} are particularly well-suited for modeling communication networks. Their ability to model complex relationships that are difficult to discover with other machine learning structures has led to their adoption in creating advanced \acp{NDT}. \cite{chang2022} presents a framework capable of discovering semantic gaps between user intents, policy generation, and the underlying network. Similarly, \cite{zhu2024} presents an \ac{NDT} architecture for 6G self-intelligent networks, including a novel method for digitally representing the global network topology based on the GraphSAGE method. \cite{boffetti2024} presents a \ac{DT} framework leveraging \ac{LSTM} for temporal estimation of \ac{QoE} based in \ac{QoS} metrics. \cite{raj2023} introduces an \ac{NDT} architecture that uses a KG for data modeling and storage, combined with a template-based method to define context. This approach enables scalable NDT management and flexible representation of network relationships.

Compared to the surveyed literature, this paper advances the state of the art in the \ac{NDT}-related research by making the following novel contributions:
\begin{itemize}
    \item Applying a multi-dimensional knowledge graph architecture to \ac{NDT}. The existing literature traditionally focuses on a single perspective, such as network state~\cite{chang2022}. The present model combines information from three different knowledge graphs: the network topology, the network state, and the application state knowledge graph. 
    \item Integrating the network state information with \ac{QoE} indicators. Existing approaches often treat network topology and service quality data as separate domains. This work establishes a unified framework where the physical network structure directly informs the estimation of \ac{QoE} for various applications. 
    \item Providing a systematic comparison and evaluation of different \ac{GNN} architectures as they are applied to \ac{NDT}. Rather than adopting a single \ac{GNN} approach, we evaluate multiple \ac{GNN}-based architectures using various performance metrics, providing insights into which architectural choices are most effective for network performance prediction.
    \item Showcasing the generalization capabilities of the proposed model, which, unlike current work, is trained and validated using a real-world dataset made available by the RIPE Atlas internet measurement system.
\end{itemize}


\section{The AI-Based Network Digital Twin Architecture}
\label{sec:architecture}

\begin{figure*}[tbp]
  \centering
  \includegraphics[width=0.73\linewidth]{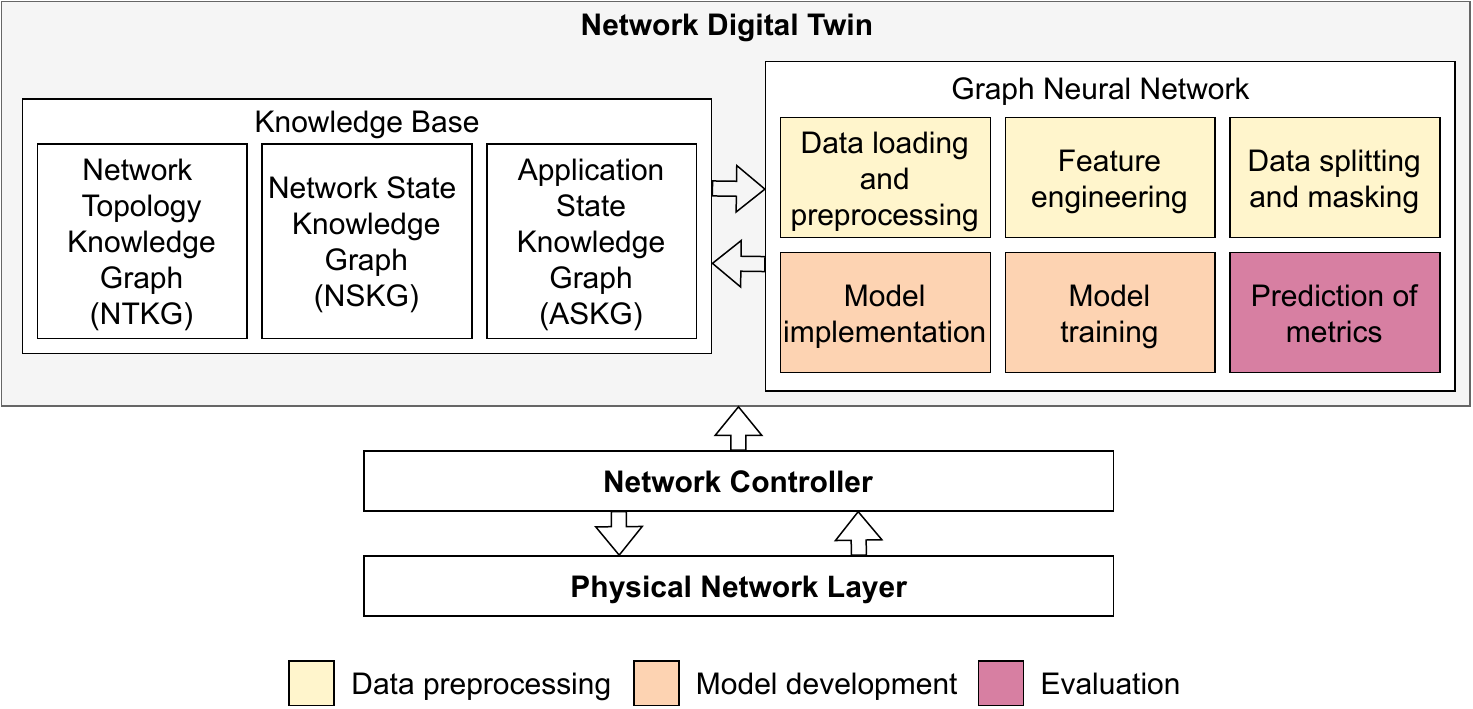}
  \caption{Simplified architecture of the system. The NDT layer on top is composed of two main components: the Knowledge Base and the Graph Neural Network, and the related processes for data retrieval.}
  \label{fig:architecture}
\end{figure*}

Our approach adopts a layered architecture, which divides the system into three primary components: the physical network layer, the network controller, and the digital twin layer~\cite{chang2022, tao2022}, as shown in~\cref{fig:architecture}. The physical network layer comprises the actual network infrastructure, including physical entities that form the end-to-end network, such as terminals (or hosts), forwarding and routing devices. The network controller is an intermediary layer that coordinates data collection from the infrastructure, allowing the \ac{NDT} to get statistics such as link state, link utilization, and logical network topology. The network digital twin layer, the focus of this work, is the core component and is responsible for maintaining a virtual replica of the network and performing predictions. 

The digital twin layer, located at the top of \cref{fig:architecture}, comprises the knowledge base and the \ac{GNN}-based prediction model. The knowledge base comprises three knowledge graphs: the \ac{NTKG}, the \ac{NSKG}, and the \ac{ASKG}. The \ac{NTKG} models the interconnections between network entities (such as routers and switches), the latitude and longitude of nodes, the degree of nodes, and also metrics derived from the original features, such as node centrality and connectivity score. The \ac{NSKG} stores the state of the network devices and links, the round-trip time, jitter, packet loss rate, and bandwidth utilization. Finally, the \ac{ASKG} stores the \ac{QoE} of multiple application types. Five types of applications were defined: web browsing, video streaming, VoIP, gaming, and file transfer applications. Each application has a unique network sensitivity profile, meaning that different weights are assigned to each of the network measurements stored in the network state knowledge graph, based on how the key performance metric impacts the \ac{QoE} of the application. \Cref{tab:applications} lists all the applications and their sensitivity profile.

\begin{table}[htbp]
\caption{Application types and sensitivity profile}
\begin{center}
\begin{tabular}{@{}lrrr@{}} \toprule
Application type & RTT weight & Loss weight & Jitter weight \\ \midrule
Web browsing     & 0.5 & 0.3 & 0.2 \\
Video Streaming  & 0.2 & 0.5 & 0.3 \\
VoIP call        & 0.3 & 0.2 & 0.5 \\
Gaming           & 0.6 & 0.2 & 0.2 \\
File transfer    & 0.1 & 0.8 & 0.1 \\ \bottomrule
\end{tabular}
\label{tab:applications}
\end{center}
\end{table}

The \ac{GNN} component is designed to process the three previously described knowledge graphs. Due to its message-passing characteristics, the \ac{GNN} can capture not only local information (data stored in the nodes or edges) but also high-level network topology relationships. Therefore, the \ac{GNN} architecture must effectively propagate information across the network topology, handle dynamic graph structures, and learn informative embeddings that support precise forecasting of key network performance metrics. To this end, the developed \ac{GNN} component operates through a methodical pipeline comprising several steps, as described below.

\subsubsection{Data loading and pre-processing}
In the data acquisition phase, the system loads real-world network measurement data from the RIPE Atlas dataset~\cite{ripe-atlas}. The dataset provides network performance metrics from probes distributed worldwide. The measurement data includes ping results, traceroute command results, and network topology data.

The measurement processing and validation phase extracts key network performance metrics, including average, minimum, and maximum \ac{RTT}, jitter, and packet loss. Invalid measurements are filtered out, and the data quality is verified again after the processing. Also, during this step, the heterogeneous measurement data is normalized and transformed into standardized performance metrics structured for effective integration with graph neural network architectures.

The processed measurements are converted into a graph structure. Network probes are converted to  nodes in the knowledge graph, and connections are established based on measurement similarity and graphic proximity.

\subsubsection{Feature engineering}
Each node representing a network probe is enriched with a fixed dimension feature vector containing network performance metrics, geographic information, and topological properties. The feature extraction process creates a 9-dimensional vector that includes the average \ac{RTT}, jitter, packet loss, ASN values, latitude and longitude, measurement counts, node degree, and neighbor counts. This feature representation includes both node properties and network connectivity patterns.

The system prepares prediction targets with a focus on two critical network performance metrics: \ac{RTT} and packet loss percentage. Specific transformations, such as the logarithmic transformation and the square root transformation, are applied to \ac{RTT} values and packet loss percentages, respectively, for two primary objectives: to address the skewed distribution and enhance model convergence. To handle data outliers, the variance in the resulting data is summarized by applying \ac{MAD}. Afterward, we scale the data with clipping to ensure stable training across various network conditions and measurement scenarios.

\subsubsection{Data splitting and masking}
This step is employed to prevent the model from memorizing the training data, thereby ensuring its ability to generalize to unseen data. The dataset is divided into three independent subsets: 60\% of the data is reserved for training, 20\% of the data is reserved for validation, and the remaining 20\% is used for testing purposes. The split is performed randomly using a uniform distribution, ensuring that each subset preserves representative network characteristics. 

To prevent data leakage among the subsets and avoid false performance estimates, a mask-based approach is employed. Each graph node in the \ac{NTKG} receives a mask, and the node is only used in one of the three phases. The masking approach enables the same graph structure to be applied in all phases (training, validation, and testing). The approach also allows for maintaining the network topology, enabling message passing among the nodes while isolating information between splits.

\subsubsection{Model implementation}
For benchmarking and comparative analysis, the developed \ac{NDT} implements multiple \ac{GNN} architectures. Each  architecture employs a different message passing mechanism, as summarized below: 

\begin{itemize}
  \item GraphSAGE (SAGE): uses neighborhood sampling and aggregation.
  \item ChebNet: employs Chebyshev polynomial approximations.
  \item ResGatedGCN: combines residual connections with gating mechanisms.
  \item GraphTransformer: applies attention mechanisms with graph-aware positional encoding.
\end{itemize}
All architectures are exposed to the same input data to ensure a fair and equitable comparison.

\subsubsection{Model training}
In this phase, the model is trained against the knowledge base data. State-of-the-art techniques, including early stopping, learning rate scheduling, and gradient clipping, are employed. Mean Squared Error (MSE) loss and the AdamW optimizer are employed for adaptive learning rates and regularization. The learning rate scheduling through ReduceLROnPlateau adjusts the learning rates based on validation performance plateaus to secure convergence while limiting the parameters from exceeding their optimal values.

In this phase, we also employ a gradient management technique called gradient clipping. Its goal is to overcome a phenomenon that commonly occurs in \acp{GNN}, namely the vanishing of gradients. The vanishing of gradients affects networks with varying topologies and connectivity patterns, potentially leading to ineffective updates of weights in early layers. Applying this strategy helps maintain the training stability even with different network configurations. 

The developed \ac{NDT} prototype, including the implementations of the four evaluated \ac{GNN} architectures, along with the evaluation framework, is publicly available at~\cite{ben-taarit2025}. The open-source release ensures full reproducibility of the experiments and facilitates collaboration within the research community.

\section{Results}
\label{sec:results}


To ensure that the proposed \ac{AI-NDT} framework is validated against realistic network conditions and, therefore, that it can be generalized to real-world applications, the dataset from RIPE Atlas~\cite{ripe-atlas} was employed. This approach bridges the gap between simulation and reality, a common limitation in many \ac{NDT} studies that rely on either synthetic data or data extracted from a controlled environment. The dataset comprises various measurement types, offering a rich and complex graph structure that enables the models to learn from. It consists of 989 nodes (\ie measurement points) and 908,752 edges (network connections) with each node having nine distinct features. In particular, the size of the dataset is crucial for assessing not only the precision of the models but also their robustness when exposed to data collected from live and complex environments.

The primary objective of this evaluation is to identify the most suitable \ac{GNN} architecture that can accurately predict network performance metrics. We focus the evaluation on two metrics that directly affect \ac{QoE} indicators: the \ac{RTT} and packet loss rate. We compare the results produced by four state-of-the-art \ac{GNN} architectures: GraphTransformer, ChebNet, ResGatedGCN, and SAGE. These architectures respond to various approaches to graph-based learning, ranging from spectral methods (ChebNet) to message-passing frameworks (SAGE, ResGatedGCN) and attention mechanisms (GraphTransformer).

\begin{figure}[b]
  \centering
  \includegraphics[width=0.86\linewidth]{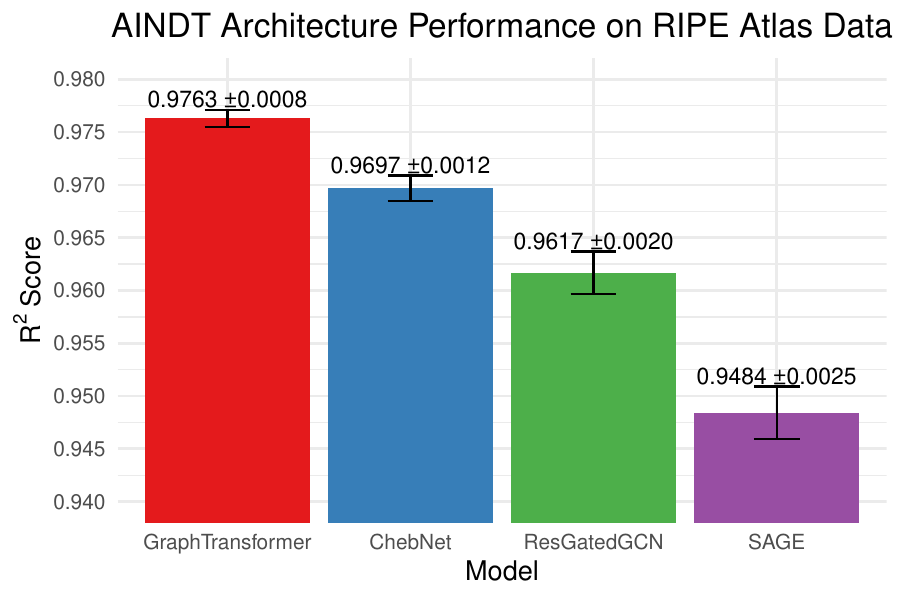}
  \caption{Coefficient of determination (\rscore) comparison among the tested models. Although SAGE presents the lowest \rscore score, its performance can still be considered very good.}
  \label{fig:plot01}
\end{figure}

\cref{fig:plot01} shows the performance of the tested models measured with the coefficient of determination (\rscore) with an over-fitting detection system. A higher \rscore value means that the model can better extract and explain the variance in the target metrics. In other words, it translates to more reliable metric predictions, which in turn translates to more dependable \ac{QoE} estimation. As illustrated in \cref{fig:plot01}, all models have achieved good results, demonstrating strong predictive capabilities, with \rscore scores exceeding 0.94. The high \rscore highlights the effectiveness of \ac{GNN} architectures for modeling complex network behaviors. 

The GraphTransformer architecture reached the highest \rscore score of 0.9763. The performance of GraphTransformer can be attributed to its attention mechanisms, which enable the identification of critical network paths and connections. It enables the model to dynamically assess the importance of different nodes and connections within the network graph. For example, the GraphTransformer can capture the information that, for a non-congested link, a change in the packet loss rate has a greater impact than a change in link utilization in predicting \ac{RTT}. For a network operator, this means the \ac{AI-NDT} can accurately predict how changes in network topology or traffic patterns will affect the \ac{QoS} for different applications, ranging from latency-sensitive gaming to bandwidth-intensive video streaming or file transfer. 

ChebNet also shows outstanding performance with an \rscore score of 0.9697, as shown in \cref{fig:plot01}. The results demonstrate that the spectral approach adopted by ChebNet can effectively capture global properties of the network. Therefore, ChebNet is a strong candidate for well-connected topologies (\ie nodes with high degree), since the model can abstract and understand system-wide performance characteristics, such as the routing of packets using multiple paths or capturing how one application can interfere with another.

\begin{figure}[b]
  \centering
  \includegraphics[width=0.855\linewidth]{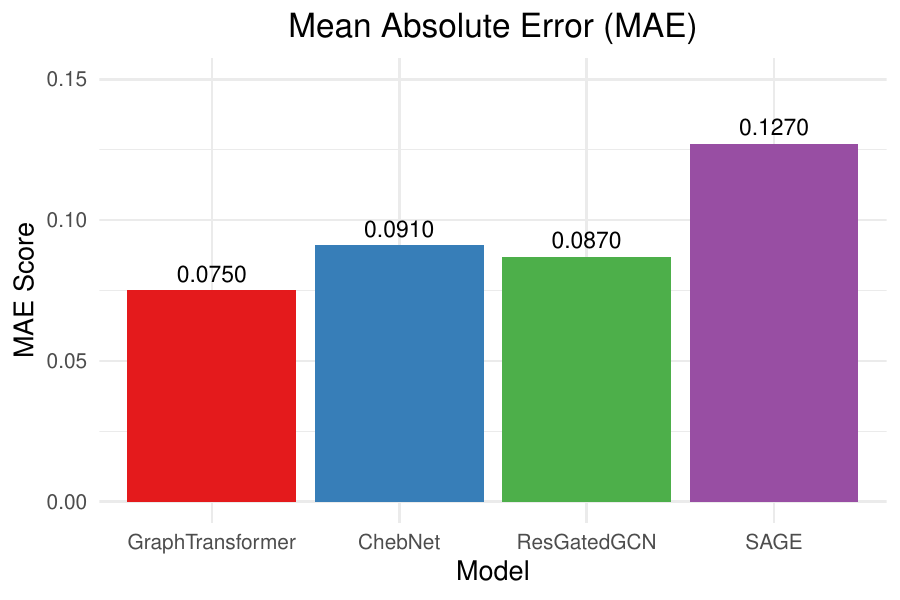}
  \caption{Mean Absolute Error (MAE) score for different GNN architectures in predicting network RTT and packet loss metrics. The MAE shows the average magnitude of error in predictions by each model.}
  \label{fig:plot02}
\end{figure}

In addition to the overall accuracy, it is crucial to analyze how often the models predict wrong values and the magnitude of the errors. For example, recurring errors with a small magnitude may not cause a significant impact for network planning tasks, since networks are usually over-provisioned. However, frequent minor errors might lead a user of a VoIP or video conference application to rate the service with a very low \ac{QoE} rating. \cref{fig:plot02} shows the \ac{MAE} for the tested \ac{GNN} architectures. The \ac{MAE} provides a measure of the average error. GraphTransformer again achieved the best result with the lowest \ac{MAE} score of 0.0750. This value indicates that the predictions of GraphTransformer are, on average, the closest to real-world values measured by the physical twin. \Acp{NDT} built on GraphTransformer architectures is particularly suitable for capacity planning tasks. 


\begin{figure}[t]
  \centering
  \includegraphics[width=0.855\linewidth]{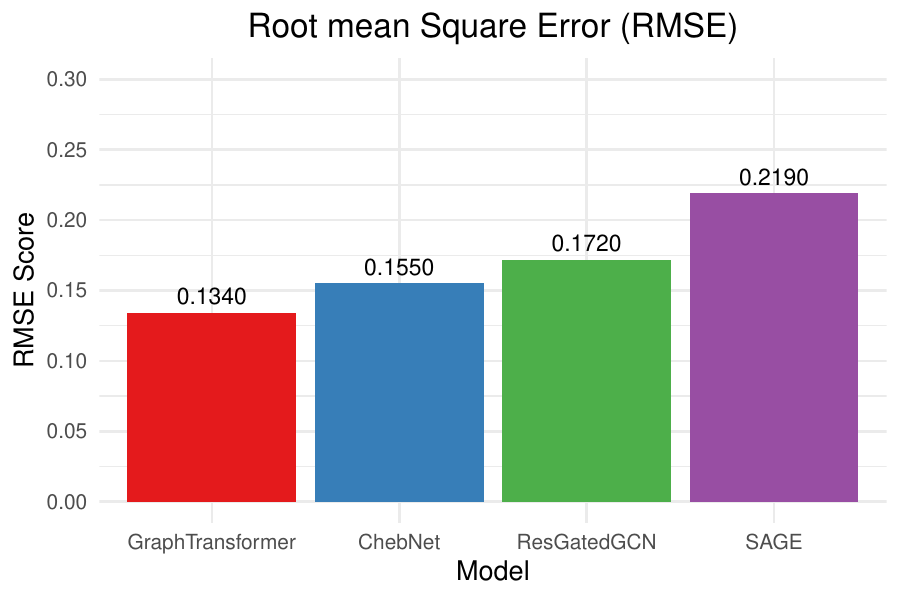}
  \caption{Root Mean Square Error (RMSE) score for the evaluated GNN models. The RMSE heavily penalizes larger prediction errors. }
  \label{fig:plot03}
\end{figure}

The \ac{RMSE} penalizes larger prediction deviations. Compared to \ac{MAE}, the \ac{RMSE} can detect errors that might not be recurrent, but have high magnitude. A low \ac{RMSE} is crucial for \ac{NDT} systems that test actions that should be instantly applied to the physical twin network. For example, an \ac{NDT} that operates in conjunction with an autonomous network management system may trigger incorrect network reconfigurations by predicting incorrect \acp{RTT} and packet loss rates, and consequently result in a low \ac{QoE} rating. \cref{fig:plot03} shows that the GraphTransformer also has the lowest RMSE, making this model a good fit for scenarios where significant errors in predicting \ac{QoE} might lead to incorrect decision-making and consequently destabilize the network.

\begin{figure}[b]
  \centering
  \includegraphics[width=0.85\linewidth]{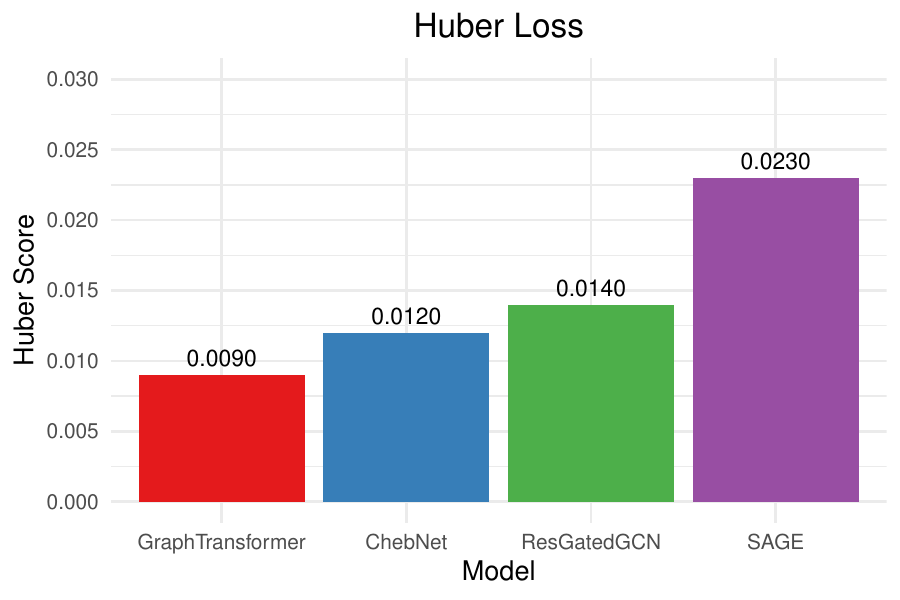}
  \caption{Huber loss for the GNN architectures in predicting network performance metrics. A lower Huber loss means that the model is less sensitive to outliers. GraphTransformer is again the model with the best results.}
  \label{fig:plot04}
\end{figure}

Network telemetry data is often noisy, either due to incorrect measurements by the telemetry tools or simply because of unexpected events in the network, such as link or interface failure. The Huber loss helps evaluate how well the models can handle those outlier measurements. The results for the Huber loss evaluation are shown in \cref{fig:plot04}, which confirms GraphTransformer's superior robustness. Its low Huber loss suggests that it is less affected by anomalous measurements.

\begin{figure}[tbp]
  \centering
  \includegraphics[width=0.85\linewidth]{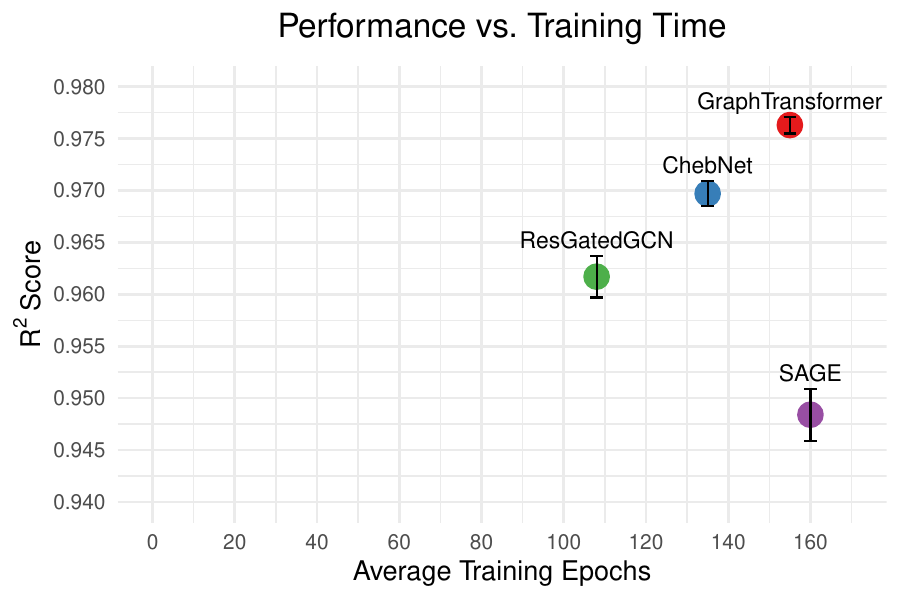}
  \caption{Trade-off between prediction performance (\rscore score) and training efficiency. GraphTransformer holds the best prediction performance, but its architecture is complex and therefore costly to train. }
  \label{fig:plot05}
\end{figure}

While the accuracy of the \ac{NDT} is critical for applications with \ac{QoE} requirements and sensitive to network conditions, the use of computational resources and training time might be vital in scenarios with limited resources. In a dynamic network environment, models may require retraining to adapt to changing conditions. For such scenarios, we analyze the tradeoff between prediction performance (\rscore score) and the training efficiency of the evaluated model measures in epochs to converge. As shown in \cref{fig:plot05}, while GraphTransformer is the top performer, it is also one of the models that takes longer to converge. Its complex attention mechanism, while powerful, is also computationally intensive. It is worth noting that ChebNet results are excellent. The model achieved the second-highest \rscore while also being the second-fastest model to convert. This balance makes it an excellent candidate for implementing an \ac{NDT} where both high accuracy and reasonable training times are desired.

\subsection{Discussions}

Our findings show that a multi-layered knowledge graph (KG) architecture combining network topology, device state, and application state proved highly effective for the implementation of \acp{NDT}. By providing the \ac{GNN} models with a holistic, contextualized view of the network, our framework enabled prediction accuracy exceeding 97\%, demonstrating that integrating these diverse data perspectives is crucial for high-fidelity \ac{QoE} estimation. This finding also shows that the adopted data structure enables the creation of \ac{NDT} that fulfill the requirements of such tools. 

The comparative analysis of the four tested models revealed that the GraphTransformer architecture is the most effective for the prediction task in \ac{NDT}. The attention mechanism in GraphTransformer is particularly well-suited for identifying critical paths and dependencies in complex, heterogeneous networks, making it the superior choice for applications that demand the highest accuracy. However, the superior performance comes at a cost of additional training epochs. On the other hand, the presented trade-off analysis shows that for use cases where rapid prototyping and deployment are paramount, the SAGE architecture provides the fastest training time while keeping accuracy at acceptance levels. 

\section{Conclusions}
\label{sec:conclusions}

This work presents an AI-based \ac{NDT} to simulate complex computer networks. The framework leverages a multi-knowledge graph architecture that combines network topologies and application state. The experiments and results show that the \ac{NDT} predictions have high accuracy. It employs advanced methodologies with statistical validation to detect and avoid overfitting, a common problem in ML-based tools. Our approach advances the state-of-the-art by integrating a multi-perspective knowledge graph architecture with a robust, comparative evaluation of advanced \ac{GNN} models trained on a real-world dataset. Additionally, the proposed architecture demonstrates its ability to scale easily, simulating networks with approximately 1000 nodes.

Although this work presents significant results and guidelines on model \ac{ML} model selection for future implementations of \ac{NDT}, we acknowledge that certain limitations exist, which open up promising avenues for future investigation. The current framework was trained and validated using a large-scale, real-world dataset obtained from the internet infrastructure. It is unclear whether the results will hold for smaller and different types of topologies, such as campus networks, \ac{IoT} networks, or data centers. Future enhancements could also include incorporating more dynamic features and integrating time-series analysis to capture temporal patterns more effectively. Finally, deploying the system in a controlled, real-time environment would be a valuable next step to validate its performance and responsiveness in a production setting.

\section*{Acknowledgment}
This work was partially supported by the HORSE (Holistic, omnipresent, resilient services for future 6G wireless and computing ecosystems) project, founded by the Smart Networks and Services Joint Undertaking (SNS JU) under the European Union’s Horizon Europe research and innovation programme under Grant Agreement No. 101096342.

\bibliographystyle{IEEEtran}
\bibliography{references}

\end{document}